\documentclass[11pt,tightenlines,eqsecnum,floats,aps,amsmath,amssymb,nofootinbib,prd,shownopacs,floatfix]{revtex4}

\usepackage{amsmath,amssymb,amsfonts,dcolumn,color,graphicx,graphics,latexsym}
\usepackage[mathscr]{eucal}
\usepackage[latin1]{inputenc}
\usepackage{enumerate}
\newcommand{\be}{\begin{equation}}
\newcommand{\ee}{\end{equation}}
\newcommand{\ba}{\begin{eqnarray}}
\newcommand{\ea}{\end{eqnarray}}

\newcommand{\bmult}{\nopagebreak[3]\begin{multline}}
\newcommand{\emult}{\end{multline}}

\begin{document}

\title{Resolution of strong singularities and geodesic completeness in loop quantum Bianchi-II spacetimes}
\author{Sahil Saini}
\email{ssaini3@lsu.edu}
\author{Parampreet Singh}
\email{psingh@phys.lsu.edu}
\affiliation{ Department of Physics and Astronomy,\\
Louisiana State University, Baton Rouge, LA 70803, U.S.A.}

\begin{abstract}
Generic resolution of singularities and geodesic completeness in the loop quantization of Bianchi-II spacetimes with arbitrary minimally coupled matter is investigated. 
Using the effective Hamiltonian approach, we examine two available quantizations: one based on the connection operator and second by treating extrinsic curvature as connection via gauge fixing.  It turns out that for the 
connection based quantization, either the inverse triad modifications or imposition of  weak energy condition is necessary to obtain a resolution of all strong singularities and geodesic completeness. 
In contrast, the extrinsic curvature based quantization generically resolves all strong curvature singularities and results in a geodesically complete effective spacetime without inverse triad modifications or energy conditions. In both the 
quantizations, weak curvature singularities can occur resulting from divergences in pressure and its derivatives at finite densities. These are harmless events beyond which geodesics can be extended. Our work generalizes previous results 
on the generic resolution of strong singularities in the loop quantization of isotropic, Bianchi-I and Kantowski-Sachs spacetimes.

\end{abstract}

\maketitle

\section{\bf Introduction}

Existence of singularities is a generic feature of general relativity (GR), a property which points out limits of its validity. This provides  impetus for the search for a more 
complete theory of gravity which includes quantum gravitational effects and hopefully results in resolution of spacetime singularities. The characterization of singularities is generally done via 
geodesic incompleteness which is the most telling manifestation of singularities. However, geodesic incompleteness may or may not be associated with a curvature pathology -- a divergence of one or more 
curvature invariants  or of the components of Riemann tensor \cite{clarke1975quasi-regular,ellis1977singular}. From the perspective of curvature pathologies, singularities can be said to be of two types - strong curvature singularities and weak curvature singularities. Strong curvature singularities or strong singularities (for short) are defined to be those that crush any in-falling object to zero volume regardless of the properties of the object \cite{ellis1977singular,tipler1977singularities,ck-1985conditions}. Clarke has shown that, given that there are no topological obstruction to geodesic extension, geodesics can always be extended except in the case of strong singularities \cite{clarke-book}. On the other hand, weak singularities are harmless. At these singularities, tidal forces are extremely strong but not strong enough to destroy an arbitrarily strong in-falling object. And geodesics can be extended beyond them in classical GR (see for eg. \cite{lazkoz,ps09}).

A measure of success of quantum theory of gravity will be whether it can resolve all strong curvature singularities. On the fundamental issue of singularity resolution, loop quantum gravity has been quite successful in the last decade. 
In loop quantum cosmology (LQC) \cite{as-status} -- a quantization of symmetry reduced spacetimes, cosmological singularities have been resolved at the full quantum level in various isotropic \cite{aps3, all-iso} and anisotropic spacetimes \cite{awe-bianchi1, awe-bianchi2, we-bianchi9, pswe, numlsu-4}, and also in polarized Gowdy models using a hybrid quantization \cite{hybrid}. In all of these investigations focusing on big bang or occasionally big crunch singularities, the classical singularity is replaced by a bounce. Evidence of bounce is also found in other frameworks based on loop quantum gravity, including group field theory \cite{gft} and quantum reduced loop gravity \cite{alesci}. Remarkably the discrete quantum geometry responsible for singularity resolution in LQC can be captured using an effective spacetime description obtained using geometric formulation of quantum mechanics \cite{vt} which has been verified using extensive numerical simulations \cite{aps3,ps12,numlsu-4,numlsu-1}. Using this effective spacetime description rigorous results have been obtained on genericity of singularity resolution in LQC. 
It has been shown that for the spatially flat isotropic and homogeneous models \cite{ps09}, Bianchi-I spacetime \cite{ps11} and Kantowski-Sachs spacetime \cite{ks-strong} with arbitrary matter all strong singularities are resolved and effective spacetime is geodesically complete. An important feature of these results is that in various LQC and loop quantized black hole spacetimes, expansion and shear scalars are universally bounded 
\cite{cs-geom,ps09,bgps-spatial,ps-proc,ps11,ck-closed2,pswe,ks-bound,cs-ks,ks-strong,cuervo-paper}, but the curvature invariants and the 
time derivative of expansion scalar can diverge \cite{wands,ps09,ps11,ks-strong,gowdy-sing}. For all the quantizations studied so far, these 
curvature invariant diverging events turn out to be harmless. They correspond to weak singularities beyond which geodesics can be extended. 
Phenomenological studies in LQC confirm this. It turns out that in the isotropic models, strong singularities of the type big bang/crunch, 
big rip and big freeze are resolved \cite{ps09,psfv,sst}, but sudden singularities do not get resolved \cite{wands,ps09}.\footnote{The big 
rip, big freeze and sudden singularities can be caused by phantom fluids (see for eg. \cite{ssd}) or for exotic equations of state.} Strong 
singularities are also resolved in loop quantization of Bianchi-I spacetime \cite{csv} and Kantowski-Sachs spacetimes 
\cite{ks-constant,cs-ks}. However, in the Bianchi-I spacetime, where there are more than one viable loop 
quantizations\footnote{The viability here implies requiring that the resulting physics is free of fiducial cell, yields infra-red limit as 
GR and a well defined scale at which Planck effects occur. For more details, see Ref. \cite{cs-unique,cs-geom}.}, the choice of the 
quantization seems to play an important role when a weak singularity is ignored or not, at least in some situations \cite{ps16}.

Our goal in this manuscript is to extend the results on generic resolution of singularities and geodesic completeness to the effective spacetime description of loop quantized Bianchi-II spacetime. 
These spacetimes are of special interest due to the findings of the Belinski-Khalatnikov-Lifshitz (BKL) conjecture and subsequent analytical and numerical investigations in GR \cite{bkl1971, Berger1998}. These studies have established a generic existence of a chaotic oscillatory behavior in the asymptotic vicinity of the classical space-like singularity. In particular, 
at every spatial point the evolution towards the singularity can be described by an infinite sequence of Kasner epochs, where each Kasner epoch is a particular Bianchi-I spacetime, also known as Mixmaster behavior. 
The above mentioned chaotic oscillations are transitions between these Kasner epochs which are well described by Bianchi-II spacetime. Bianchi-I spacetime has been shown to be free of all strong curvature  
singularities in effective dynamics of LQC \cite{ps11}. It is therefore natural to generalize these results to Bianchi-II spacetimes to gain insights on the 
way quantum gravitational effects result in singularity resolution in generic scenarios. 

The generalization of results of the loop quantization of Bianchi-I spacetime to Bianchi-II spacetime is not straightforward. It has been already noted that the presence of spatial 
curvature term makes things rather non-trivial with respect to the boundedness of energy density, expansion and shear scalars \cite{bgps-spatial}. A highlight of such a non-triviality 
is also found in the loop quantization of Kantowski-Sachs spacetime where above results were recently generalized \cite{ks-strong}. The situation in loop quantization of Bianchi-II spacetime is more complex. 
Here to express field strength in terms of holonomies over closed loops yielding an algebra of almost periodic functions is not possible. Instead one has to consider holonomies over open 
loop sections and assign the area using the underlying homogeneity of space. Unlike the other loop quantized spacetimes there is a departure in the strategy. Using inputs from loop quantum 
gravity, one can introduce a connection operator written in terms of holonomies and use that to express the classical 
Hamiltonian constraint in the quantum form. This results in what we will refer to as connection operator based loop quantization or for short `A' quantization because the operator analog of 
Ashtekar-Barbero connection $A^i_a$ is used as a key object in this construction. Though the resulting quantization leads to that of Bianchi-I spacetime in the limit when spatial curvature vanishes, 
physics of this loop quantized spacetime is quite unexpected. Unlike other loop quantum spacetimes, the modifications to classical constraint originating from holonomies in the field strength part of the 
Hamiltonian constraint {\it{are not sufficient}} 
to universally bind expansion and shear scalars. It was found that one has to impose weak energy condition to achieve the same results as were obtained for the loop quantized Bianchi-I spacetime \cite{bgps-spatial}. 
A more troubling situation exists for 
a generalization of `A' quantization to Bianchi-IX spacetime where expansion and shear scalars may even diverge even after imposing weak energy condition. 
A deeper understanding of these issues led to revisiting the so called `K' quantization in LQC \cite{pswe}, first used more than a decade ago in Refs. \cite{kevin-no-bound,date,kevin-thesis}. 
Note that in the homogeneous spacetimes with a spatial curvature, while quantizing it is possible to perform a gauge fixing such that the extrinsic curvature $K^i_a$ can be treated as a connection. 
One can then use this `connection' for parallel transport along open loop segments to obtain quantum Hamiltonian constraint \cite{kevin-no-bound,date,kevin-thesis}. The resulting quantum Hamiltonian constraint is similar to the one in `A' quantization but has subtle important differences. In particular, for the loop quantized Bianchi-IX spacetime it yields universal bounds on expansion and shear scalars without imposition of any extra conditions as in `A' quantization \cite{pswe}. In a sense, the `K' quantization is a natural generalization of loop quantization of isotropic and Bianchi-I spacetimes. 

This work, as previous studies on this issue, is based on effective spacetime description of LQC whose validity has been tested rigorously for isotropic \cite{numlsu-1} and recently for anisotropic spacetimes \cite{numlsu-4}. 
Assuming the validity of effective dynamics, we show that for the `A' quantization if one only takes into account holonomy modifications arising from the field strength part in the quantum Hamiltonian constraint then 
the strong singularities are not resolved. The effective spacetime is geodesically incomplete. This is unlike the situation in loop quantization of isotropic, Bianchi-I and Kantwoski-Sachs spacetimes and the 
cause of this lies in potential singularities arising due to one of the triads vanishing in a finite time evolution, and unboundedness of associated directional Hubble rate. To remedy this situation there are two avenues. First, is to impose weak energy condition. This results in a lower bound on the energy density which translates to the troubling behavior of one of the triads and directional Hubble rates getting cured. The second alternative is to include the inverse triad modifications in the quantum Hamiltonian constraint. These are on many occasions ignored in discussion of LQC because they seem to be not as significant as holonomy modifications,\footnote{The reason is tied to the general qualitative feature of dynamics in isotropic LQC. Large macroscopic universes at late times bounce at Planck curvature but at volumes much greater than Planck. The inverse triad modifications become important only near Planck volume.} but are the second main inputs of quantum geometry in the loop quantization procedure. But for the spatial curved models, such as Bianchi-II spacetime, their importance in singularity resolution has been emphasized strongly \cite{bgps-spatial,ck-closed2,corichi-bianchi3}. The main result from the `A' quantization of the Bianchi-II spacetime is that all strong singularities are resolved and effective spacetime is geodesically complete in the effective spacetime if inverse triad modifications are included. If these are excluded, then weak energy condition is necessary to be imposed. Curvature invariants can still diverge if either of these conditions are imposed, but such divergences turn out to be weak singularities. Our work highlights the importance of not ignoring the inverse triad modifications. If one does, there is price to pay in `A' quantization. Either incorporate energy conditions, which so far have never been needed for singularity resolution in LQC, or give up generic resolution of singularities. 

In a striking contrast, the `K' quantization yields results on generic resolution of singularities without any  extra conditions. 
All the triads turns out to be finite and bounded in finite time evolution. The directional Hubble rates, expansion and shear scalars are universally bounded. The effective spacetime is geodesically complete and there are no strong singularities in a finite time evolution. There can be divergences in curvature invariants and the 
time derivative of expansion scalar, but such divergences are weak singularities. To emphasize, there is no need to impose either any energy condition or include inverse triad modifications to achieve generic resolution of singularities in the `K' quantization of Bianchi-II spacetime.

This paper is organized as follows. In Sec. II, we provide a brief overview of the Hamiltonian formulation in terms of Ashtekar variables of the classical Bianchi-II spacetime. 
Key dynamical equations, expansion and shear scalars, and curvature invariants are discussed in this section. In Sec. III, we describe the `A' quantization of the Bianchi-II model in LQC. 
We consider the effective dynamics of the `A' quantization without incorporating the inverse triad corrections in Sec. IIIA and describe various features including the proof that 
one of the triads can vanish in finite time. We then include the weak energy condition and show the boundedness of all the triads in finite time evolution. The inverse triad modifications 
for the `A' quantization and the resulting dynamical features are discussed in Sec. IIIB. The results turn out to be similar as in Sec. IIIA with weak energy condition imposed.  
In Sec. IV we discuss the effective dynamics for the `K' quantization of Bianchi-II spacetime. The boundedness of energy density, expansion and shear scalars is proved without 
introducing any energy condition or inverse triad modification. Both the `A' quantization, including inverse triad modifications or the weak energy condition, and the `K' quantization 
do not forbid events where components of Riemann tensor, curvature invariants and time derivative expansion scalar diverge. This is proved in Sec. V. In Sec. VI, we show that 
effective spacetime in `A' quantization including either inverse triad modifications or weak energy condition is geodesically complete. 
If either of the two are absent, then geodesics can break down in finite time evolution. In contrast, the effective spacetime is geodesically complete unconditionally in `K' quantization. 
The effective spacetime is maximally extendible. We discuss the strength of singularities in Sec. VII. We show that there are no strong singularities in the effective spacetime spacetime of `K' quantization or `A' quantization, with either inverse triad modifications or weak energy condition imposed, in finite time evolution. Potential curvature invariant divergent events in these quantizations turn out to be weak singularities. This is not the case for `A' quantization in the absence of weak energy condition or the inverse triad modifications. In that case, strong singularities can occur in finite time evolution.  We summarize the results in Sec. VIII.

\section{Classical dynamics of diagonal Bianchi-II spacetime}
In this section, we start with a summary of the basics of Bianchi-II model in the Hamiltonian framework in the connection and triad variables, obtain Hamilton's equations from the classical Hamiltonian and list important scalars needed in the analysis.  Bianchi-II spacetime is a homogeneous and anisotropic spacetime, whose any given spatial slice has a set of three independent Killing vector fields ($\zeta_1,\zeta_2$ and $\zeta_3$, say) along which the metric is preserved. These Killing vector fields form a closed algebra: $[\zeta_i,\zeta_j]=C^k_{ij}\zeta_k$ with 
the only non-zero structure constants as $C^1_{23}=-C^1_{32} \in \mathbb{R}^{+}$ for the Bianchi-II spacetime. We consider the spatial topology to be $\mathbb{R}^3$. As a result, introduction of a fiducial cell is necessary to define a symplectic structure. Let us denote $V_o$ as the fiducial volume of a cuboid fiducial cell with fiducial lengths $L_i$ $(i = 1..3)$
 with fiducial metric ${\mathring{q}}_{ab}$ on the hypersurfaces. The algebra for the fiducial triads $\mathring e_{i}^{a}$ compatible with the fiducial metric can be worked out from the Killing fields. The spacetime metric for the Bianchi-II model can then be put in a simplified form as follows 
\be
\mathrm{d} s^2= - \mathrm{d} t^2 + a_1^2(\mathrm{d} x - \frac{L_1}{L_2 L_3}\alpha z \mathrm{d} y)^2 + a_2^2 \mathrm{d}y^2 + a_3^2 \mathrm{d}z^2 \label{metric}
\ee
where the constant $\alpha = C^1_{23}$. 

In loop quantum gravity, the classical gravitational phase space variables are the matrix valued Ashtekar-Barbero connection $A_{a}^{i}$ and the conjugate triad $E_{i}^{a}$. Due to homogeneity, these variables are symmetry-reduced and are given as follows,
\begin{equation}
E_{i}^{a}=\tilde{p_i}\sqrt{|\mathring q|} {\mathring e}_{i}^{a} \quad \text{and} \quad A_{a}^{i}=\tilde{c}^i {\mathring\omega}_{a}^{i}
\end{equation}
where ${\mathring\omega}_{a}^{i}$ are the fiducial co-triads. The symmetry reduced connection and triad components satisfy: 
\begin{equation}
\left\lbrace \tilde{c}^i,\tilde{p_j} \right\rbrace = \frac{8\pi G \gamma}{V_o} \delta_{j}^{i} 
\end{equation}
where $\gamma$ is the Barbero-Immirzi parameter. Since the symplectic structure depends on the fiducial cell parameters, we absorb them in the redefinition of the reduced triad and connection variables as below,
\begin{equation}
p_i = \frac{V_o \tilde{p_i}}{L_i} \quad \text{and} \quad c^i = L_i \tilde{c}^i ~.
\end{equation}
So that the non-zero Poisson brackets become:
\begin{equation}
\left\lbrace c^i,p_j \right\rbrace =8\pi G \gamma \delta_{j}^{i} .\label{poisson1}
\end{equation}

The directional scale factors in the metric are related to the triads as:
\begin{equation}
a_i = \frac{\sqrt{|p_1 p_2 p_3|}}{L_i p_i} ~.
\end{equation}
The connection components are related to the time derivatives of the scale factors using Hamilton's equations which can be obtained from the classical Hamiltonian. In terms of the symmetry reduced connection and triad variables, it can be written as \cite{awe-bianchi2}:
\begin{eqnarray}
\mathcal{H}_{cl} &=& -\frac{1}{8 \pi G\gamma^2 \sqrt{|p_1 p_2 p_3|}}\bigg[p_1 p_2 c_1 c_2 + p_2 p_3 c_2 c_3 + p_3 p_1 c_3 c_1 + \alpha \epsilon p_2 p_3 c_1 \nonumber \\
& & - (1 + \gamma^2)\bigg(\frac{\alpha p_2 p_3}{2 p_1}\bigg)^2 \bigg] + \rho \sqrt{|p_1 p_2 p_3|} ~. \label{hamiltonian_classical}
\end{eqnarray}
Here $\rho$ denotes the energy density of a minimally coupled matter field. It is defined as $\rho = {\cal H}_m/V$ where ${\cal H}_m$ denotes matter Hamiltonian and $V = V_o |a_1 a_2 a_3|$ denotes the physical 
volume of the fiducial cell. The constant $\epsilon = \mathrm{sgn}(p_1) \mathrm{sgn}(p_2) \mathrm{sgn}(p_3)$ is either $+1$ or $-1$ depending on whether the triads are right-handed or left-handed respectively. 
Notice that each term in the above Hamiltonian is left invariant under change in the orientation in any of the physical triads because the corresponding $p_i$ and $c_i$ both flip the sign under such a transformation. 
Due to this reason, making a transformation in the triad space from one octant to another leaves the Hamiltonian invariant. Hence, the RHS of \eqref{hamiltonian_classical} only depends on the magnitude of the triad and connection variables. 
We can find out an expression for the energy density on the constraint surface from the vanishing of the Hamiltonian constraint ${\cal C}_{cl} = 8 \pi G {\cal H}_{cl} \approx 0$. It is to be noted that the value of energy density depends only the magnitude of the triad and connection variables, not on the triad orientation, and which in the positive octant is:
\begin{equation}
\rho=\frac{1}{8 \pi G\gamma^2}\bigg[ \frac{c_3 c_1}{p_2} + \frac{c_1 c_2}{p_3} +  \frac{c_2 c_3 +\alpha c_1}{p_1} - (1 + \gamma^2)\alpha^2 \frac{p_2 p_3}{4p_1^3}\bigg] ~.
\end{equation}
The energy density can diverge in various scenarios, e.g. when either of the connection variables diverge, or when the triad variables vanish.

To find out the time evolution of the scale factors, the Hubble rates, the expansion and shear scalars and curvature invariants, we are going to need the equations of motion. Using the Hamiltonian and the Poisson brackets \eqref{poisson1}, the equations of motion for the triad variables can be written as,
\begin{align}
\dot p_1 & = -8\pi G \gamma \frac{\partial\mathcal{H}_{cl}}{\partial c_1} = \frac{p_1}{\gamma \sqrt{|p_1 p_2 p_3|}}\bigg( p_2 c_2 + p_3 c_3 + \frac{\alpha \epsilon p_2 p_3}{p_1} \bigg),\label{eom-triad-classical1}\\
\dot p_2 & = -8\pi G \gamma \frac{\partial\mathcal{H}_{cl}}{\partial c_2} = \frac{p_2}{\gamma \sqrt{|p_1 p_2 p_3|}}\bigg( p_3 c_3 + p_1 c_1 \bigg),  \label{eom-triad-classical2}\\
\intertext{and}
\dot p_3 & = -8\pi G \gamma \frac{\partial\mathcal{H}_{cl}}{\partial c_3} = \frac{p_3}{\gamma \sqrt{|p_1 p_2 p_3|}}\bigg( p_1 c_1 + p_2 c_2 \bigg) . \label{eom-triad-classical3}
\end{align}
These equations seem to suggest that the triad variables can diverge under certain conditions, e.g. when the connection variables diverge. The equations also do not seem to prevent the triad variables from vanishing at some point during evolution. We will see that these are precisely the places where one or more quantities of interest such as energy density, Hubble rates or the curvature invariants show pathological behavior.

Let us calculate the Hubble rates in terms of the phase space variables,
\begin{equation}
H_i=\frac{\dot a_i}{a_i}=\frac{1}{2}\bigg(\frac{\dot p_j}{p_j}+\frac{\dot p_k}{p_k}-\frac{\dot p_i}{p_i} \bigg) \quad \text{where} \quad i \neq j \neq k ; \quad i,j,k \in \lbrace 1,2,3 \rbrace \label{hubble}
\end{equation}
The Hubble rates diverge whenever the ratios $\frac{\dot p_i}{p_i}$ diverge and these possibilities are not excluded in the classical evolution of the triad variables as seen from eqs.(\ref{eom-triad-classical1}-\ref{eom-triad-classical3}).

We further calculate the expansion scalar, its time derivative and shear scalar. They describe the effects of the spacetime on an extended object as it travels along a geodesic, and are crucial in the analysis of singularities as they capture the Raychaudhuri equation. For co-moving observers the expansion scalar ($\theta$), its time derivative and shear scalar ($\sigma^2$) are given by,
\begin{eqnarray}
\theta &=& \frac{\dot V}{V} = H_1+H_2+H_3 = \frac{1}{2} \sum_{i=1}^{3} \frac{\dot p_i}{p_i} \label{exxscalar}\\
\dot \theta &=& \frac{1}{2} \sum_{i=1}^{3} \bigg(\frac{\ddot p_i}{p_i} - \bigg(\frac{\dot p_i}{p_i} \bigg)^2 \bigg) \label{thetadot} \\
\sigma^2 &=& \sum_{i=1}^{3} (H_i-\theta)^2 = \frac{1}{3} \bigg((H_1-H_2)^2+(H_2-H_3)^2+(H_3-H_1)^2 \bigg) ~. \label{shscalar}
\end{eqnarray}

It is important to notice from \eqref{hubble}, \eqref{exxscalar} and \eqref{shscalar} that the behavior of expansion and shear scalar depend only on the behavior of quantities $\frac{\dot p_i}{p_i}$. On the other hand the time derivative of expansion scalar also depends on the accelerations $\ddot p_i$. This distinction will turn out to have important consequences in LQC in the boundedness of $\theta$ and $\sigma^2$ and 
potential divergences in $\dot \theta$. 

Finally we come to the curvature invariants such as the Ricci and Kretschmann scalar. Both such scalars are obtained from the second derivatives of the metric with respect to the coordinates. Given that the metric depends on the time coordinate through the triad variables, we see that these curvature invariants can at most depend on the second time derivatives of the triad variables. For example, the Ricci scalar comes out to be,
\begin{equation}
R=\frac{\ddot p_1}{p_1}+\frac{\ddot p_2}{p_2}+\frac{\ddot p_3}{p_3}-\alpha^2 \frac{p_2 p_3}{2p_1^3} \label{ricciscalar}
\end{equation}
and the Kretschmann scalar in terms of triads and its derivatives becomes
\begin{eqnarray}
K&=&\frac{1}{4p_1^6 p_2^4 p_3^4} \bigg[11 \alpha^4 p_2^6 p_3^6 + 14 p_1^6 p_3^4 \dot p_2^4-2\alpha^2 p_1 p_2^5 p_3^3 \bigg(29 p_3^2 \dot p_1^2 - 26 p_1 p_3 \dot p_1 \dot p_3 + 5 p_1^2 \dot p_3^2 \bigg) \nonumber \\
&& -4p_1^5 p_2 p_3^2 \dot p_2^2 \bigg(4 \dot p_2(p_3 \dot p_1 + p_1 \dot p_3) + 5 p_1 p_3 \ddot p_2 \bigg) + 4 p_1^4 p_2^2 p_3^2 \bigg(p_1^2 \dot p_2^2 \dot p_3^2 \nonumber \\
&& + p_3^2 \lbrace \dot p_2^2 \left[\dot p_1^2 + 3 p_1 \ddot p_1 \right] + 2 p_1 \dot p_1 \dot p_2 \ddot p_2 + 3 p_1^2 \ddot p_2^2 \rbrace + p_1 p_3 \dot p_2 \lbrace 2 \dot p_3 \left[ 2 \dot p_1 \dot p_2 + p_1 \ddot p_2 \right] + 3 p_1 \dot p_2 \ddot p_3 \rbrace \bigg) \nonumber \\
&& + 2 p_1^3 p_2^3 p_3 \bigg( -5 \alpha^2 p_3^4 \dot p_2^2 - 8 p_1^3 \dot p_2 \dot p_3^3 + p_3^3 \lbrace 4 \dot p_1 \dot p_2 \left[-2 \dot p_1^2 + p_1 \ddot p_1 \right] + 2 p_1 \left[ 3 \dot p_1^2 - 2 p_1 \ddot p_1 \right] \ddot p_2 \rbrace \nonumber \\
&& + 2 p_1^2 p_3 \dot p_3 \lbrace \dot p_3 \left[ 4 \dot p_1 \dot p_2 + 3 p_1 \ddot p_2 \right] + 2 p_1 \dot p_2 \ddot p_3 \rbrace + 4 p_1 p_3^2 \lbrace \dot p_3 \left[ \dot p_2 ( 2 \dot p_1^2 - 3 p_1 \ddot p_1) - 3 p_1 \dot p_1 \ddot p_2 \right] \nonumber \\
&& - p_1 \left[ 3 \dot p_1 \dot p_2 + p_1 \ddot p_2 \right] \ddot p_3 \rbrace \bigg) + 2 p_1^2 p_2^4 \bigg( 26 \alpha^2 p_3^5 \dot p_1 \dot p_2 + 7 p_1^4 \dot p_3^4 + p_3^4 [ 7 \dot p_1^4 - 14 \alpha^2 p_1 \dot p_2 \dot p_3 - 10 p_1 \dot p_1^2 \ddot p_1 \nonumber \\
&& + 6 p_1^2 \ddot p_1^2 ] - 2 p_1^3 p_3 \dot p_3^2 \left[ 4 \dot p_1 \dot p_3 + 5 p_1 \ddot p_3 \right] + 2 p_1 p_3^3 [ 2 \dot p_1 \dot p_3 ( -2 \dot p_1^2 + p_1 \ddot p_1) \nonumber \\
&& p_1 ( 3 \dot p_1^2 - 2 p_1 \ddot p_1) \ddot p_3 ] + 2 p_1^2 p_3^2 \left[ \dot p_3^2 ( \dot p_1^2 + 3 p_1 \ddot p_1) + 2 p_1 \dot p_1 \dot p_3 \ddot p_3 + 3 p_1^2 \ddot p_3^2 \right] \bigg) \bigg] ~. \label{kretschmann}
\end{eqnarray}
Any term in the expression of the Ricci or the Kretschmann scalar contains upto second time derivatives of the triads. This is also true of any components of the Riemann tensor, or any other tensor obtained from second derivatives of the metric such as the Weyl tensor. It turns out that for our analysis we do not need to consider higher order invariants. As we will see in Section VII that we only need to look at the properties of Riemann tensor components to rule out strong singularities. 

\section{Effective dynamics: `A' Quantization}

In this section we study the effective dynamics of the connection operator based `A' quantization. This quantization was first introduced in Ref. 
\cite{awe-bianchi1,awe-bianchi2}, and a different factor ordering which results in a decoupling of ths singularity at the vanishing of the triads was proposed in 
Ref. \cite{hybrid}.  In this quantization a connection operator is constructed using holonomies over open edges to express the classical 
Hamiltonian constraint in the quantum form.  Here we note that the effective dynamics has been shown be extremely accurate in capturing the 
underlying quantum dynamics  for a large variety of states including for the anisotropic spacetimes \cite{numlsu-4}. In the quantum 
Hamiltonian constraint, modifications from quantum geometry enter via two avenues. First by expressing classical connection as quantum 
operators using holonomies over open edges, and second by expressing inverse triads in terms of holonomies. The latter modifications ensure 
that the eigenvalues of inverse triad operators do not diverge even when the eigenvalues of triad operator vanish. As mentioned 
earlier, in the previous studies of analysis of geodesic completeness and strong singularities in LQC in isotropic \citep{ps09}, Bianchi-I 
\cite{ps11} and Kantowski-Sachs models \cite{ks-strong}, the desired results could be obtained without including inverse triad 
corrections. However, in the case of loop quantization of Bianchi-II and Bianchi-IX in `A' quantization, it was suspected 
that the inverse-triad corrections may turn out to be important for singularity resolution \cite{bgps-spatial}. Indeed this turns out to be 
the case as we will demonstrate in this section. 
In the following subsection we  first consider the effective dynamics of the `A' quantization without inverse triad 
modifications of Bianchi-II spacetime and analyze the behavior of the triad variables and their derivatives.   We then note the difficulties posed when we do not include the inverse triad corrections in the effective Hamiltonian. Namely that one of the triads can vanish in finite time evolution and one of the directional Hubble rates is 
unbounded.  Imposition of weak energy 
condition becomes necessary and results in a finite and bounded behavior of triads and directional Hubble rates and expansion and shear 
scalars. As we will see later, this proves to be an important input for generic resolution of strong singularities. The same results are 
obtained in Sec. IIIB without imposing weak energy condition but incorporating inverse triad modifications in the effective 
Hamiltonian. 

For the purpose of our analysis, we  restrict ourselves to the positive octant in the triad space after introducing some of the main dynamical equations in the following subsection. The analysis can be similarly performed for the negative octant too without changing any result. Here we note that in the quantization proposed in \cite{awe-bianchi2}, restricting oneself to the positive octant in the triad space implies choosing those quantum states which are symmetric under reflections in the triad space. Restricting to one octant can also be justified by choosing a different factor ordering in the Hamiltonian constraint, as done in \cite{hybrid} for Bianchi-I model, which ensures that different octants are decoupled. The effective Hamiltonian then obtained is the same in both the cases, either following Ref. \cite{awe-bianchi2} and restricting to symmetric states and positive octant (as originally done to study effective dynamics in Ref. \cite{awe-bianchi2}) or considering the effective dynamics following from extension of analysis in Ref. \cite{hybrid} to Bianchi-II spacetime. This is also discussed in remarks 1 and 2 in the following subsection. There is a subtlety with the effective dynamics corresponding to Ref. \cite{hybrid} and the decoupling of the singularity which is elaborated in remark 3.

\subsection{`A' quantization without inverse triad corrections}

As mentioned in the discussion above, the effective Hamiltonain turns out to be the same for different ways of factor ordering in the quantum Hamiltonian (either Ref. \cite{awe-bianchi2} or \cite{hybrid}) once the restriction to positive octant is made. However, let us start by following the construction in Ref. \cite{awe-bianchi2} and write the effective Hamiltonian for the connection based quantization choosing lapse $N=1$ for arbitrary minimally coupled matter as follows without making any assumption yet on the orientation:
\begin{eqnarray}
\mathcal{H} &=& -\frac{\sqrt{|p_1 p_2 p_3|}}{8 \pi G\gamma^2 \Delta l_{pl}^2}\bigg[\mathrm{sgn}(p_1) \mathrm{sgn}(p_2) \sin(\bar\mu_1 c_1)\sin(\bar\mu_2 c_2)+ \mathrm{cyclic} \bigg]\nonumber \\
&& -\frac{1}{8 \pi G\gamma^2}\bigg[\mathrm{sgn(p_1)} \frac{\alpha |p_2 p_3|}{(\sqrt{\Delta}l_{pl}) |p_1|} \sin(\bar\mu_1 c_1) - (1+\gamma^2)\frac{{\alpha}^2 |p_2 p_3|^{3/2}}{4 {|p_1|}^{5/2}} \bigg]+\rho\sqrt{|p_1 p_2 p_3|} ~. \label{HamiltonianA}
\end{eqnarray}
Here $\bar \mu_i$ label the lengths of the loops used to express the connection operator in terms of holonomies. These are determined by the underlying quantum geometry as 
\be
\bar\mu_1=\lambda \sqrt{\frac{|p_1|}{|p_2 p_3|}} \quad \bar\mu_2=\lambda \sqrt{\frac{|p_2|}{|p_1 p_3|}} \quad \bar\mu_3=\lambda \sqrt{\frac{|p_3|}{|p_1 p_2|}} \label{mubar}
\ee
where $\lambda^2 = \Delta l_{pl}^2 = 4\sqrt{3}\pi \gamma l_{\mathrm{Pl}}^2$ is the minimum non-zero area of the loop in loop quantum gravity.

The resulting Hamilton's equations for triads are:
\begin{align}
\frac{\dot p_1}{p_1} &= -\frac{8\pi G \gamma}{p_1} \frac{\partial\mathcal{H}}{\partial c_1} = \frac{1}{\gamma \lambda}  (\mathrm{sgn}(p_2) \sin(\bar\mu_2 c_2)+\mathrm{sgn(p_3)} \sin(\bar\mu_3 c_3)+ \lambda \xi)\cos(\bar\mu_1 c_1),  \label{p1dotA} \\
\frac{\dot p_2}{p_2} &= -\frac{8\pi G \gamma}{p_2} \frac{\partial\mathcal{H}}{\partial c_2} = \frac{1}{\gamma \lambda}  (\mathrm{sgn(p_3)} \sin(\bar\mu_3 c_3)+\mathrm{sgn(p_1)} \sin(\bar\mu_1 c_1))\cos(\bar\mu_2 c_2), \label{p2dotA} \\
\intertext{and}
\frac{\dot p_3}{p_3} &= -\frac{8\pi G \gamma}{p_3} \frac{\partial\mathcal{H}}{\partial c_3} = \frac{1}{\gamma \lambda}  \mathrm{(sgn(p_1)} \sin(\bar\mu_1 c_1)+\mathrm{sgn(p_2)} \sin(\bar\mu_2 c_2))\cos(\bar\mu_3 c_3) .\label{p3dotA} \\
\end{align}
where 
\begin{equation}
\xi=\alpha \sqrt{\frac{|p_2 p_3|}{|p_1|^3}} ~\label{xi}
\end{equation}

We now investigate whether the triads can vanish or become infinite in a finite time evolution. 
Let $t_0 $ be some time in the present at which $p_1, p_2, p_3$ have some given finite values $p_1^0, p_2^0, p_3^0$. Then from \eqref{p2dotA} we have
\begin{equation}
\int_{p_2^0}^{p_2(t)}\frac{dp_2}{p_2}= \int_{t_0}^t \frac{1}{\gamma \lambda} (\mathrm{sgn(p_3)} \sin(\bar\mu_3 c_3)+\mathrm{sgn(p_1)} \sin(\bar\mu_1 c_1)) dt . \label{p_2-1A}
\end{equation}
Upon formal integration, we get
\begin{equation}
p_2(t) = p_2^0  \exp\left\lbrace\frac{1}{\gamma \lambda} \int_{t_0}^t\bigg((\mathrm{sgn(p_3)} \sin(\bar\mu_3 c_3)+\mathrm{sgn(p_1)} \sin(\bar\mu_1 c_1))\cos(\bar\mu_2 c_2)\bigg) dt\right\rbrace . \label{p_2-2A}
\end{equation}
\\

It is important to note that \eqref{p_2-2A} implies that $p_2(t)$ has the same sign as $p_2^o$ as long as the equation does not break down (e.g. when a singularity is encountered in finite time evolution). Using a similar argument, we can say that this is also true for $p_1(t)$ and $p_3(t)$, i.e. the sign of $p_i$ does not change as far as the evolution is non-singular.
\\

\noindent
\textbf{Remark 1:} Note that the effective Hamiltonian and the equations of motion have $\mathrm{sgn(p_i)}$ factors in various terms. We can clean up further calculations if we could restrict the $p_i$'s to a particular octant in the triad space. In this regard, we can follow the approach in \cite{awe-bianchi1} and \cite{awe-bianchi2} where at the quantum level, the basis states for the kinematical Hilbert space are chosen to be symmetric under reflections in the triad space. 
The effective Hamiltonian then can be obtained by restriction to a positive octant \cite{awe-bianchi1,awe-bianchi2}. Following this, at the effective level we can choose our initial conditions such that the triads start out in the positive octant in our analysis. From the discussion in the previous paragraph, we then find that the  triads remain in the same octant where their initial conditions were specified during a non-singular time evolution.
\\

\noindent
\textbf{Remark 2:} As an alternative to above procedure, we can take the approach of Ref. \cite{hybrid} where a suitable choice of factor ordering of the quantum Hamiltonian constraint leads to a decoupling of different octants. The states corresponding to singularity at vanishing of the triads is also decoupled from the other states. In that case we can choose to focus on the positive octant and we are ensured that the triads will remain in the positive octant during the non-singular time evolution. Note that unlike the quantization in Ref. \cite{awe-bianchi2}, no restriction is required for the symmetry of the states. The assumption of a positive octant is therefore more natural when effective Hamiltonian is assumed to arise from quantization of Bianchi-II spacetime based on Ref. \cite{hybrid}. \\


In the following analysis in this manuscript, we assume that one of the above mentioned choices is adopted so that we can restrict our attention to the positive octant in the triad space. This allows us to drop all the $\mathrm{sgn(p_i)}$ factors and remove the modulus signs in the effectoive Hamiltonian, and the initial values $p_i^o$ are assumed to be positive definite. \\

Now we return to equation \eqref{p_2-2A}, where we now drop all the $\mathrm{sgn}(p_i)$ factors. In equation \eqref{p_2-2A}, since $ | (\sin(\bar\mu_3 c_3)+\sin(\bar\mu_1 c_1))\cos(\bar\mu_2 c_2)|\leq 2$, the integration (inside the exponential) over a finite time is finite. Hence, for any finite time past or future evolution, we have: 
\begin{equation}
0<p_2(t)<\infty \label{p_2-3A}
\end{equation}
From \eqref{p3dotA}, we can conclude similarly that $0<p_3(t)<\infty$. However, the equation for $p_1$ has an extra term containing $\xi$. Upon formal integration of \eqref{p1dotA}, we can write
\begin{equation}
p_1(t) = p_1^0  \exp\left\lbrace\frac{1}{\gamma \lambda} \int_{t_0}^t\bigg((\sin(\bar\mu_2 c_2)+\sin(\bar\mu_3 c_3))\cos(\bar\mu_1 c_1)\bigg)dt \right\rbrace \exp\left\lbrace\frac{\alpha}{\gamma} \int_{t_0}^t \sqrt{\frac{p_2 p_3}{p_1^3}} \cos(\bar\mu_1 c_1) dt \right\rbrace \label{p1A}
\end{equation}
where we have explicitly substituted the expression for $\xi$. The first exponential term remains finite and non-zero for all finite times since $ |(\sin(\bar\mu_2 c_2)+\sin(\bar\mu_3 c_3))\cos(\bar\mu_1 c_1)|\leq 2$. However, the same can not be said about the second exponential term in the above expression.

Let us note that potential singularities can occur when $p_1$ either vanishes or diverges in a finite time evolution. Let us further investigate whether the second exponential in \eqref{p1A} can either vanish or diverge. If $p_1$ starts from a finite non-zero value and diverges in a finite time, then the LHS of \eqref{p1A} diverges. However, since $p_2$ and $p_3$ are finite for finite times, the integral in the second exponential of \eqref{p1A} stays finite. This means that the RHS of \eqref{p1A} remains finite. This shows that $p_1$ going to infinity is not consistent with eq. \eqref{p1A}.

However, vanishing of $p_1$ is consistent with equation \eqref{p1A}, if the term $\cos(\bar\mu_1 c_1)$ happens to be negative in the region when $p_1$ is approaching zero. This can lead to a potential cigar type singularity. To see this, let us write down the triads in terms of the directional scale factors, along with the conditions we just obtained for triad variables from equations of motion in the effective loop quantum spacetime.
\begin{align}
p_1 & \propto |a_2 a_3|, \quad p_1<\infty, \nonumber \\
p_2 & \propto |a_3 a_1|, \quad 0<p_2<\infty \nonumber \\
\intertext{and}
p_3 & \propto |a_1 a_2|, \quad 0<p_3<\infty. \nonumber
\end{align}

We see that $p_1$ can go to zero if $a_1$ diverges and $a_2, a_3$ go to zero at the same rate. Such an approach to the singularity is a cigar singularity. Note that the expansion and shear scalars are obtained from first derivatives of the metric, hence depend on quantities of the type $\frac{\dot p_i}{p_i}f(p_1,p_2,p_3)$. Thus, when $p_1$ goes to zero, the expansion and shear scalars also diverge.
\\

\noindent
\textbf{Remark 3:} Above we found that for the `A' quantization, it is possible that $p_1$ vanishes in a finite time evolution obtained from the effective Hamiltonian which is assumed to be valid at all the scales. 
If the effective dynamics is assumed to be obtained from Ref. \cite{hybrid} then this result points to a limitation of the 
effective dynamics. Recall that in remark 2 we discussed that for the quantization prescription of Ref. \cite{hybrid}, the singularity is decoupled from the evolution. However, the corresponding effective Hamiltonian does not include the information 
about this decoupling. The reason is tied to the way effective Hamiltonian is derived in LQC, where an underlying approximation is to consider volumes greater than the Planck volume \cite{vt}. Further, extensive numerical simulations for 
anisotropic models have shown that effective dynamics becomes less reliable for states which probe deep Planck volumes \cite{numlsu-4}. 
Thus, using the effective dynamics of Bianchi-II spacetime, which is assumed to be valid at all scales, we find a singularity at vanishing of triad $p_1$. Such a singularity in the effective dynamics will be absent in the quantum evolution of Bianchi-II spacetime with physical states using quantization prescription in Ref. \cite{hybrid}. For numerical simulations with such states, we expect to see significant departures between quantum evolution and effective dynamics.  \\


We next turn our attention to the energy density. From the vanishing of the effective Hamiltonian constraint corresponding to the effective Hamiltonian \eqref{HamiltonianA}, we see that the energy density is dynamically equal to,
\begin{equation}
\rho = \frac{1}{8 \pi G\gamma^2 \lambda^2} \bigg[(\sin(\bar\mu_1 c_1)\sin(\bar\mu_2 c_2)+ \text{cyclic terms}\bigg] +\frac{1}{8 \pi G\gamma^2}\bigg[ \frac{\xi}{\lambda} \sin(\bar\mu_1 c_1) - (1-\gamma^2)\frac{\xi^2}{4} \bigg] ~. \label{rhoA}
\end{equation}
It can be shown that the expression for energy density has a global maxima at $\xi=\frac{2}{\lambda (1+\gamma^2)}=\xi_0$ and $\sin(\bar\mu_1 c_1)=1$ \cite{awe-bianchi2}. However, even though the energy density has an upper bound, it becomes negative and diverges at the above mentioned cigar singularities, along with the divergences in the expansion and shear scalars \cite{bgps-spatial}. As we will discuss in Sec. VII, at these cigar singularities the tidal forces 
become infinite causing them strong curvature type singularities with breakdown of geodesic evolution. We thus have here a first ever result from the loop quantization of a homogeneous spacetime where strong singularities can occur.

A remedy from such potential singularities was hinted in Ref. \cite{bgps-spatial}. The idea is to impose weak energy condition, the physical assumption that energy density should remain non-negative. 
The latter with the fact that energy density is bounded from above, allows us to restrict $p_1$ to non-zero values as we proceed to show in the remaining part of this subsection.



Eq. \eqref{rhoA} implies that dynamically the energy density is an inverse parabola if considered as a function of $\xi$ and becomes negative for large deviation in $\xi$ away from $\xi_0$ in either direction \cite{bgps-spatial}. If we impose the weak energy condition on the matter content to keep the energy density non-negative, then $\xi$ is bounded. Since both $p_2$ and $p_3$ are finite for any finite time, weak energy condition implies that $p_1$ remains non-zero for all of the finite time. We have already argued that the equations of motion imply that $p_1$ cannot diverge either. Together with  the results on $p_2$ and $p_3$ proved earlier, the equations of motion together with the weak energy condition on the matter content gives the following generic result that for all finite times
\begin{equation}
0<p_i<\infty \quad \text{and} \quad 0<\frac{\dot p_i}{p_i}<\infty . \label{piboundedA}
\end{equation}
This in turn implies that the scale factors, the volume, the Hubble rates, the expansion and shear scalars all remain non-zero and finite for all finite times. However the curvature invariants such as the Ricci and Kretschmann scalar depend on higher time derivatives of the triad variables, which may still diverge. But we will see in the section VI and VII that this still implies geodesic completeness and resolves all strong cosmological singularities in Bianchi-II spacetime. 

To summarize the results so far, in the connection based `A' quantization of the Bianchi-II spacetime strong singularities and geodesic incompleteness persists unless we impose weak energy condition. Note that this result was derived in the absence of inverse triad modifications in the Hamiltonian. In the next subsection we will see that we do not need to assume any energy conditions if we include the inverse triad corrections in the effective Hamiltonian of the `A' quantization.

\subsection{`A' quantization with inverse triad corrections}

As discussed earlier, loop quantum gravity effects can manifest themselves in two forms in the effective dynamics - holonomy corrections and inverse triad corrections. Although the Hamiltonian \eqref{HamiltonianA} of the previous subsection  incorporates the holonomy corrections, it still lacks the inverse triad corrections. There are several terms in the classical Hamiltonian that contain inverse powers of the triad $p_1$. To include these inverse triad corrections in our effective Hamiltonian in all the terms, we consider the eigenvalues of the operator $\widehat{{\vert p_1 \vert}^{-1/4}}$ as obtained 
in Ref. \cite{awe-bianchi2}. The action of  $\widehat{{\vert p_1 \vert}^{-1/4}}$ is given by
\begin{equation}
\widehat{{\vert p_1 \vert}^{-1/4}} \vert p_1 p_2 p_3 \rangle = g(p_1) \vert p_1 p_2 p_3 \rangle
\end{equation}
where
\begin{equation}
g(p_1)=\frac{(p_2 p_3)^{1/4}}{\sqrt{2\pi \gamma \sqrt{\Delta} l_{pl}^3}} ( \sqrt{\vert v+1 \vert} - \sqrt{\vert v-1 \vert}); \quad \quad v=\frac{\sqrt{p_1 p_2 p_3}}{2\pi \gamma \sqrt{\Delta} l_{\mathrm{Pl}}^3} ~.
\end{equation}

The effective Hamiltonian with inverse triad corrections for the lapse $N=1$ then becomes,
\begin{eqnarray}
\mathcal{H} &=& -\frac{\sqrt{p_1 p_2 p_3}}{8 \pi G\gamma^2 \Delta l_{pl}^2}\bigg[(\sin(\bar\mu_1 c_1)\sin(\bar\mu_2 c_2)+\sin(\bar\mu_2 c_2)\sin(\bar\mu_3 c_3)+\sin(\bar\mu_3 c_3)\sin(\bar\mu_1 c_1))\bigg]\nonumber \\
&& -\frac{1}{8 \pi G\gamma^2}\bigg[ \frac{\alpha p_2 p_3}{(\sqrt{\Delta}l_{pl})}g(p_1)^4 \sin(\bar\mu_1 c_1) - (1-\gamma^2)\frac{{\alpha}^2 (p_2 p_3)^{3/2}}{4}g(p_1)^{10} \bigg]+\rho\sqrt{p_1 p_2 p_3} \label{HamiltonianAT}
\end{eqnarray}
In comparison to the effective Hamiltonian \eqref{HamiltonianA}, the above Hamiltonian does not contain any inverse powers of the triads. The Hamilton's equations for triads then become:
\begin{align}
\frac{\dot p_1}{p_1} &= -\frac{8\pi G \gamma}{p_1} \frac{\partial\mathcal{H}}{\partial c_1} = \frac{1}{\gamma \lambda}  \bigg(\sin(\bar\mu_2 c_2)+\sin(\bar\mu_3 c_3)+ \lambda \alpha \sqrt{p_2 p_3} \frac{g(p_1)^4}{\sqrt{p_1}} \bigg)\cos(\bar\mu_1 c_1), \label{p1dotAT} \\
\frac{\dot p_2}{p_2} &= -\frac{8\pi G \gamma}{p_2} \frac{\partial\mathcal{H}}{\partial c_2} = \frac{1}{\gamma \lambda}  (\sin(\bar\mu_3 c_3)+\sin(\bar\mu_1 c_1))\cos(\bar\mu_2 c_2), \\
\intertext{and}
\frac{\dot p_3}{p_3} &= -\frac{8\pi G \gamma}{p_3} \frac{\partial\mathcal{H}}{\partial c_3} = \frac{1}{\gamma \lambda}  (\sin(\bar\mu_1 c_1)+\sin(\bar\mu_2 c_2))\cos(\bar\mu_3 c_3) .
\end{align}

Let us now examine whether triads $p_i$ ever vanish or diverge in the finite time evolution. We note that the equations for $p_2$ and $p_3$ are unchanged compared to the case when inverse triads are absent. The arguments presented in equations \eqref{p_2-1A}, \eqref{p_2-2A} and \eqref{p_2-3A} of Sec. IIIA hold here too, so we can conclude, in analogy, that $p_2$ and $p_3$ will remain positive-definite and finite for all finite time evolution if we start with finite and positive definite initial data for the triad variables.

However the time evolution equation for $p_1$ is now different from the one in Sec. IIIA due to inverse triad corrections. It turns out that it makes a critical difference for the generic resolution of singularities. If we start from initial data $p_1^o$ where $0<p_1^o<\infty$, then eq. \eqref{p1dotAT} can be integrated to yield:
\begin{equation}
p_1(t) = p_1^0  \exp\left\lbrace\frac{1}{\gamma \lambda} \int_{t_0}^t\bigg((\sin(\bar\mu_2 c_2)+\sin(\bar\mu_3 c_3))\cos(\bar\mu_1 c_1)\bigg)dt \right\rbrace \exp\left\lbrace\frac{\alpha}{\gamma} \int_{t_0}^t \sqrt{p_2 p_3} \frac{g(p_1)^4}{\sqrt{p_1}} \cos(\bar\mu_1 c_1) dt \right\rbrace \label{p1AT}
\end{equation}
The first exponential term in the above equation remains finite and positive definite as the integrand in the integral in that term is bounded for all times. The second exponential term decides whether $p_1$ can either diverge or vanish. However, it follows from the expression for the function $g(p_1)$ that the factor $\frac{g(p_1)^4}{\sqrt{p_1}}$ tends to zero if $p_1$ either goes to zero or diverges. Since $p_2$ and $p_3$ are both finite valued for all finite times, this implies that the second exponential term also remains finite in these limits. Hence both of the scenarios, either $p_1$ going to zero or diverging, are inconsistent with the eq. \eqref{p1AT} and are thus excluded from the dynamics. Hence, we conclude that $p_1$, like $p_2$ and $p_3$, also remains positive definite and finite for all times.

Hence, the inclusion of inverse triad corrections in the effective Hamiltonian means that effective dynamics is such that for all finite times
\begin{equation}
0<p_i<\infty \quad \text{and} \quad 0<\frac{\dot p_i}{p_i}<\infty . \label{piboundedAT}
\end{equation}

This result is in striking contrast with the situation in the previous subsection where we needed to impose weak energy condition to obtain this crucial result. Eq. \eqref{piboundedAT} in turn implies that the volume, the Hubble rates, and the expansion and shear scalar remain positive definite and finite for all finite time evolution. Finally, let us look at the energy density in this case. Using the vanishing of the corresponding Hamiltonian constraint, we obtain
\begin{eqnarray}
\rho &=& \frac{1}{8 \pi G\gamma^2 \Delta l_{pl}^2}\bigg[(\sin(\bar\mu_1 c_1)\sin(\bar\mu_2 c_2)+\sin(\bar\mu_2 c_2)\sin(\bar\mu_3 c_3)+\sin(\bar\mu_3 c_3)\sin(\bar\mu_1 c_1))\bigg]\nonumber \\
&& + \frac{1}{8 \pi G\gamma^2 \sqrt{p_1 p_2 p_3}}\bigg[ \frac{\alpha p_2 p_3}{(\sqrt{\Delta}l_{pl})}g(p_1)^4 \sin(\bar\mu_1 c_1) - (1-\gamma^2)\frac{{\alpha}^2 (p_2 p_3)^{3/2}}{4}g(p_1)^{10} \bigg] ~.
\end{eqnarray}
Due to the result \eqref{piboundedAT}, all the terms in the expression for energy density are well behaved for all finite times. Hence the energy density remain finite for all finite times.

We see that the results which are crucial for resolution of singularities, which in the previous subsection were obtained by imposing weak energy condition, are implied by Hamilton's equations themselves if we incorporate the inverse triad corrections. So we either have to include the inverse triad corrections, or impose weak energy condition in order to resolve the singularities in the `A' quantization. In the next section we demonstrate that in the case of extrinsic curvature based `K' quantization, these results are obtained without having to include inverse triad corrections or imposition of weak energy condition.

 \section{Effective dynamics: `K' Quantization}

In order to overcome the problem of non-availability of algebra of almost  periodic functions of connection when considering holonomies 
over closed loops, a problem which also plagues loop quantization of Bianchi-II spacetime,  an idea proposed before `A' quantization was to 
consider holonomies of extrinsic curvature over open edges \cite{kevin-no-bound,date,kevin-thesis}. Note that the extrinsic curvature can 
be 
treated as a connection via a gauge fixing in the Gauss constraint. This procedure results in an inequivalent quantization -- the `K' 
quantization -- of the Hamiltonian constraint than the `A' quantization. This quantization is known as the `K' quantization 
\cite{kevin-thesis}, and has been used in different models of LQC, most recently in the context of Bianchi-IX spacetime where it leads to 
bounded behavior of expansion and shear scalars, a feature absent in `A' quantization \cite{pswe}. It was shown that the `K' quantization 
results in physics which is qualitatively very close to the so far loop quantization models using holonomies over closed loops \cite{pswe}. 
In the following we will follow the strategy used in Ref. \cite{pswe} and study the effective dynamics of the `K' quantization of the 
Bianchi-II spacetime, and we will restrict ourselves to the positive octant in the triad space as per our discussion in the previous section. 

We start with writing the classical Hamiltonian constraint \eqref{hamiltonian_classical} in terms of components of extrinsic curvature and 
triads. Let us note that the extrinsic curvature in the Bianchi-II spacetime is given by:
\begin{equation}
K^{i}_{a} = \gamma^{-1}(A^{i}_{a}-\Gamma^{i}_{a}), \quad \text{where} \quad K^{i}_{a}=\frac{k^{i}}{L_{i}} (^o\omega^{i}_{a})
\end{equation}
where $\Gamma^{i}_{a}$ is the spin connection. To express the classical Hamiltonian constraint in terms of $k_i$ and $p_i$ we need the 
relationship between $k_i$'s  and $c_i$'s. These can be obtained from above  by plugging in the expressions 
for the terms on the RHS, or we can use the Hamilton's equations \eqref{hamiltonian_classical} to find $k_i$'s using the following relation 
:
\begin{equation}
k_i = L_i \dot{a_i} ~.
\end{equation}
The expressions for the connection components in the classical theory turn out to be:
\begin{align}
c_1 &= \gamma k_1 + \frac{\alpha \epsilon}{2} \frac{p_2 p_3}{p_1^2},  \\
c_2 &= \gamma k_2 - \frac{\alpha \epsilon}{2} \frac{p_3}{p_1}, \\
\intertext{and}
c_3 &= \gamma k_3 - \frac{\alpha \epsilon}{2} \frac{p_2}{p_1} ~. 
\end{align}
The $k_i$ and $p_i$ satisfy the following Poisson bracket: 
\begin{equation}
\left\lbrace k_i,p_j \right\rbrace=8 \pi G \delta_{ij} ~.
\end{equation}

The classical Hamiltonian (for arbitrary minimally coupled matter) \eqref{hamiltonian_classical} can then be written as 
\begin{eqnarray}
\mathcal{H}_{cl} &=& -\frac{1}{8 \pi G\gamma^2 \sqrt{p_1 p_2 p_3}}\bigg[\gamma^2 p_1 p_2 k_1 k_2 + \gamma^2 p_2 p_3 k_2 k_3 + \gamma^2 p_3 
p_1 k_3 k_1 \nonumber \\
& & - \frac{\alpha^2 \gamma^2}{4}\bigg(\frac{p_2 p_3}{p_1}\bigg)^2 \bigg] + \rho \sqrt{p_1 p_2 p_3} . \label{hamiltonian_classical_k}
\end{eqnarray}
where we have used the fact that $\epsilon^2=1$. The corresponding Hamiltonian constraint can be quantized the same way as in the 
Bianchi-IX model \cite{pswe}. The resulting effective Hamiltonian (in absence of inverse triad modifications) is given by 
\begin{eqnarray}
\mathcal{H} &=& -\frac{\sqrt{p_1 p_2 p_3}}{8 \pi G\gamma^2 \lambda^2}\bigg[\sin({\bar{\mu}}_1\gamma k_1) \sin({\bar{\mu}}_2\gamma k_2) + 
\sin({\bar{\mu}}_2\gamma k_2) \sin({\bar{\mu}}_3\gamma k_3) + \sin({\bar{\mu}}_3\gamma k_3) \sin({\bar{\mu}}_1\gamma k_1) \nonumber \\
& & - \frac{\alpha^2 \lambda^2 \gamma^2}{4} \frac{p_2 p_3}{p_1^3} \bigg] ~ + ~ \rho \sqrt{p_1 p_2 p_3} . \label{hamiltonian_k}
\end{eqnarray}
The Hamilton's equations for the triads then can be obtained as:
\begin{align}
\frac{\dot p_1}{p_1} &={\gamma \lambda}\bigg(\sin({\bar{\mu}}_2\gamma k_2) + \sin({\bar{\mu}}_3\gamma k_3)\bigg) 
\cos({\bar{\mu}}_1\gamma k_1),  \label{p1dotK} \\
\frac{\dot p_2}{p_2} &={\gamma \lambda}\bigg(\sin({\bar{\mu}}_3\gamma k_3) + \sin({\bar{\mu}}_1\gamma k_1)\bigg) 
\cos({\bar{\mu}}_2\gamma k_2),  \label{p2dotK} \\
\intertext{and}
\frac{\dot p_3}{p_3} &={\gamma \lambda}\bigg(\sin({\bar{\mu}}_1\gamma k_1) + \sin({\bar{\mu}}_2\gamma k_2)\bigg) 
\cos({\bar{\mu}}_3\gamma k_3).  \label{p3dotK}
\end{align}
We thus find that quantities $\frac{\dot p_i}{p_i}$ are bounded functions of 
time, i.e. they are bounded for all time. This implies that the Hubble rates \eqref{hubble}, the expansion scalar \eqref{exxscalar} and the 
shear scalar \eqref{shscalar} are bounded for all times. This is a considerable improvement over `A' quantization where to obtain the same 
result weak energy condition  was needed if inverse triad corrections were not included.

It can be easily shown  that the triads $p_i$ remain non-zero and finite for all finite time evolution if we start the evolution from 
some non-zero and finite initial values $p_i^o$. For example we can integrate the equation for $p_1$ to obtain:
\begin{equation}
p_1(t) = p_1^0  \exp\left\lbrace\frac{1}{\gamma \lambda} \int_{t_0}^t\bigg((\sin({\bar{\mu}}_2\gamma k_2) + \sin({\bar{\mu}}_3\gamma k_3)) 
\cos({\bar{\mu}}_1\gamma k_1)\bigg) dt\right\rbrace \label{p_1K}
\end{equation}
Since the integral inside the exponential has a bounded integrand, $p_1$ remains positive definite and finite for all finite time 
evolution, 
i.e. $0<p_1<\infty$. Similar results are obtained for the triad components $p_2$ and $p_3$. 

Let us obtain the expression for energy density. From the vanishing of the Hamiltonian constraint we obtain:
\begin{eqnarray}
\rho &=& \frac{1}{8 \pi G\gamma^2 \lambda^2}\bigg[\sin({\bar{\mu}}_1\gamma k_1) \sin({\bar{\mu}}_2\gamma k_2) + \sin({\bar{\mu}}_2\gamma 
k_2) \sin({\bar{\mu}}_3\gamma k_3) + \sin({\bar{\mu}}_3\gamma k_3) \sin({\bar{\mu}}_1\gamma k_1) \nonumber \\
& & - \frac{\alpha^2 \lambda^2 \gamma^2}{4} \frac{p_2 p_3}{p_1^3} \bigg] . \label{rhoK}
\end{eqnarray}
Since triads remain finite and bounded for all finite times, the energy density never diverges in a finite time evolution. 

Thus, in the `K' quantization one obtains in a straightforward way the boundedness of energy density, expansion and shear scalars. The 
triads never vanish in finite time evolution. Note that these results are obtained without using weak energy condition or inverse triad 
modifications. The contrast with the `A' quantization discussed in the previous section is striking. In the case of `A' quantization, these 
results can only be obtained either by incorporating inverse triad corrections or the weak energy condition. This shows the advantage of 
the `K' quantization over `A' quantization. We proceed in the next section to discuss pathologies that may still be lurking in 
curvature invariants.

\section{Potential divergences in Curvature Invariants and $\dot \theta$}

We have shown in previous sections that in the effective dynamics of LQC, in the connection operator based `A' quantization the energy density, the expansion and shear scalars are only bounded if 
we impose either weak energy conditions or include inverse triad modifications. The same conditions are needed for avoiding volume to vanish at a finite time. Whereas, in the extrinsic curvature based `K' quantization, energy density, expansion and shear scalars are bounded universally and volume never vanishes in finite time evolution. It is important to note that the latter is true without imposition of 
any weak energy condition or using inverse triad modifications. 

However, to understand the generic resolution of strong singularities and geodesic completeness we need to consider the behavior of curvature invariants and the derivative of the expansion scalar $\dot \theta$. 
The latter quantity is important to understand the focusing of geodesics. 

We first note that the Riemann tensor is obtained from the second derivatives of the metric with respect to the coordinates. Hence, the expressions for the components of the Riemann tensor contain second derivatives of the triad variables with respect to time. So the second derivatives $\ddot p_i$ will also be present in the expressions for the curvature invariants obtained from the Riemann tensor (and invariants obtained from other such tensors obtained from second derivatives of the metric) such as the Ricci and the Kretschmann scalar. As an example we have provided the expression for the Ricci and Kretschmann scalars in Sec. II (see \eqref{ricciscalar} and \eqref{kretschmann}), which depend on the ratios $\frac{\ddot p_i}{p_i}$.

In the sections on `A' and `K'- quantizations, we have shown that it is possible to obtain the result that the ratios $\frac{\dot p_i}{p_i}$ remain bounded and that the triads $p_i$ remain positive definite and finite for all finite time evolution. This means that the divergence properties of the curvature invariants such as the Ricci and Kretschmann scalar depend on the properties of the second time derivatives of the triads $\ddot p_i$, and if these diverge, so do the Ricci and Kretschmann scalar and all other higher order curvature invariants that may depend on third or higher time derivatives of the triads. Hence, in order to understand any potential divergences in the curvature invariants we need to understand the behavior of $\ddot p_i$.

Consider the expressions for the first time derivatives of the triads as given in \eqref{p1dotA} and \eqref{p1dotAT} for `A'-quantization, and \eqref{p1dotK} for `K'-quantization. They imply that the expressions for the second time derivatives of the triads, $\ddot p_i$, are going to have the first time derivatives of the connection variables, $\dot c_i$, in them. These are obtained from Hamilton's equations,
\begin{equation}
\dot c_i= 8\pi G \gamma \frac{\partial\mathcal{H}}{\partial p_i}
\end{equation}
and contain the derivatives of the energy density with respect to the triad variables $\frac{\partial \rho}{\partial p_i}$. It means that $\ddot p_i$ and consequently the curvature invariants will diverge whenever these derivatives of the energy density with respect to the triad variables $\frac{\partial \rho}{\partial p_i}$ diverge. Hence, in both `A' and `K' quantizations, the curvature invariants can diverge if the triad-derivatives of the energy density diverge at some point. If, for some specific choices of the matter content, the derivatives $\frac{\partial \rho}{\partial p_i}$ diverge at finite values of volume, energy density, and expansion and shear scalar, then we are sure to have divergences of curvature invariants.

As an illustration, consider the case of `A' quantization without the inverse triad corrections. Taking the time derivatives of equations \eqref{p1dotA}, \eqref{p2dotA} and \eqref{p3dotA}, we get the following expressions for second derivatives of the triads:
\begin{eqnarray}
\frac{\ddot p_1}{p_1} &=& \bigg(\frac{\dot p_1}{p_1} \bigg)^2 + \frac{1}{\gamma \lambda}\bigg[( {\dot {\bar{\mu}}}_2 c_2 + {\bar{\mu}}_2 \dot c_2)\cos({\bar{\mu}}_2 c_2) + ( {\dot {\bar{\mu}}}_3 c_3 + {\bar{\mu}}_3 \dot c_3)\cos({\bar{\mu}}_3 c_3) \bigg] \cos({\bar{\mu}}_1 c_1) \nonumber \\
& & - \frac{1}{\gamma \lambda}\bigg[\sin({\bar{\mu}}_2 c_2) + \sin({\bar{\mu}}_3 c_3) + \lambda \xi \bigg]( {\dot {\bar{\mu}}}_1 c_1 + {\bar{\mu}}_1 \dot c_1)\sin({\bar{\mu}}_1 c_1) + \frac{\dot \xi}{\gamma} \cos({\bar{\mu}}_1 c_1) \\
\frac{\ddot p_2}{p_2} &=& \bigg(\frac{\dot p_2}{p_2} \bigg)^2 + \frac{1}{\gamma \lambda}\bigg[( {\dot {\bar{\mu}}}_1 c_1 + {\bar{\mu}}_1 \dot c_1)\cos({\bar{\mu}}_1 c_1) + ( {\dot {\bar{\mu}}}_3 c_3 + {\bar{\mu}}_3 \dot c_3)\cos({\bar{\mu}}_3 c_3) \bigg] \cos({\bar{\mu}}_2 c_2) \nonumber \\
& & - \frac{1}{\gamma \lambda}\bigg[\sin({\bar{\mu}}_1 c_1) + \sin({\bar{\mu}}_3 c_3) \bigg]( {\dot {\bar{\mu}}}_2 c_2 + {\bar{\mu}}_2 \dot c_2)\sin({\bar{\mu}}_2 c_2) \\
\frac{\ddot p_3}{p_3} &=& \bigg(\frac{\dot p_3}{p_3} \bigg)^2 + \frac{1}{\gamma \lambda}\bigg[( {\dot {\bar{\mu}}}_1 c_1 + {\bar{\mu}}_1 \dot c_1)\cos({\bar{\mu}}_1 c_1) + ( {\dot {\bar{\mu}}}_2 c_2 + {\bar{\mu}}_2 \dot c_2)\cos({\bar{\mu}}_2 c_2) \bigg] \cos({\bar{\mu}}_3 c_3) \nonumber \\
& & - \frac{1}{\gamma \lambda}\bigg[\sin({\bar{\mu}}_1 c_1) + \sin({\bar{\mu}}_2 c_2) \bigg]( {\dot {\bar{\mu}}}_3 c_3 + {\bar{\mu}}_3 \dot c_3)\sin({\bar{\mu}}_3 c_3)
\end{eqnarray}

We see that the expression for $\ddot p_1$ contains terms that depend on $\xi$ and $\dot \xi$ (where $\xi$ is given by \eqref{xi}). Both $\xi$ and $\dot \xi$ are bounded by virtue of conditions of the type \eqref{piboundedA}, which hold in case of `A'-quantization provided either weak energy conditions are assumed or inverse triad corrections are included, and hold in case of `K'-quantization without any extra conditions. Hence the divergence properties of $\frac{\ddot p_i}{p_i}$ ultimately only depend on quantities of type $( {\dot {\bar{\mu}}}_i c_i + {\bar{\mu}}_i \dot c_i)$. Let us consider the behavior of $( {\dot {\bar{\mu}}}_1 c_1 + {\bar{\mu}}_1 \dot c_1)$ as an example:
\begin{eqnarray}
{\dot {\bar{\mu}}}_1 c_1 + {\bar{\mu}}_1 \dot c_1 &=& \frac{1}{2 \gamma \lambda} \bigg[ -(\sin(\bar\mu_1 c_1)\sin(\bar\mu_2 c_2)+ \mathrm{cyclic} ~\mathrm{terms}) + 2 \lambda \xi \sin(\bar{\mu}_1 c_1) - \frac{5 \lambda^2 (1-\gamma^2)}{4} \xi^2 \nonumber \\
& & + 8 \pi G \gamma^2 \lambda^2 \rho + (\bar{\mu}_2 c_2 - \bar{\mu}_1 c_1) (\sin(\bar\mu_3 c_3)+\sin(\bar\mu_1 c_1))\cos(\bar\mu_2 c_2)\nonumber \\
& & +  (\bar{\mu}_3 c_3 - \bar{\mu}_1 c_1) (\sin(\bar\mu_1 c_1)+\sin(\bar\mu_2 c_2))\cos(\bar\mu_3 c_3)  \bigg] + \lambda p_1 \frac{\partial \rho}{\partial p_1} ~.
\end{eqnarray}
In the above expression, note that except for $\frac{\partial \rho}{\partial p_1}$ and factors of type $(\bar{\mu}_2 c_2 - \bar{\mu}_1 c_1)$ and $(\bar{\mu}_3 c_3 - \bar{\mu}_1 c_1)$, all other factors and terms are well behaved if weak energy condition is assumed. Let us look further at the properties of the factors of type $(\bar{\mu}_2 c_2 - \bar{\mu}_1 c_1)$. Using \eqref{mubar} we note that,
\begin{equation}
\bar{\mu}_2 c_2 - \bar{\mu}_1 c_1 = \frac{\lambda}{V}(c_2 p_2 - c_1 p_1) ~.
\end{equation}
Similarly expressions are obtained for similar combinations of $\bar \mu_j c_j$ and $\bar \mu_i c_i$ for various values of $i$ and $j$. Let us look specifically at $(c_2 p_2 - c_1 p_1)$. Its derivative can be written as follows,
\begin{equation}
\frac{d}{dt}(c_2 p_2 - c_1 p_1)=V \bigg(p_2 \frac{\partial \rho}{\partial p_2} - p_1 \frac{\partial \rho}{\partial p_1} \bigg) + \textrm{ other  bounded  terms} ~.
\end{equation}
Therefore, the fate of the term ultimately depends on the derivatives of the energy density $\frac{\partial \rho}{\partial p_i}$ as we had foreseen in our discussion at the beginning of this section. 

Repeating this exercise for the `A' quantization with inverse triad corrections and `K' quantization, one reaches the same result. Therefore, we  conclude that in all the quantizations discussed in this paper, divergences in quantities $\frac{\partial \rho}{\partial p_i}$ result in divergences in curvature invariants. We are going to call these divergences as ``pressure divergences" because in case of matter with vanishing anisotropic stress, $\frac{\partial \rho}{\partial p_i}$ are proportional to the pressure and they diverge if the pressure diverges.

The fate of the $\dot \theta$ is the same as the curvature invariants. The time derivative of the expansion scalar as given in eq. \eqref{thetadot} also depends on the second time derivatives of the triads. It follows from our discussion above that $\dot \theta$ will also diverge at these "pressure divergences" due to the divergences in the quantities $\frac{\partial \rho}{\partial p_i}$. However, as we have noted already in previous sections that the expansion scalar itself remains finite for all finite times by virtue of the effective dynamics in 'K' quantization or in the case of 'A' quantization with either weak energy condition imposed or with inverse triad corrections. Thus, we note that at these ``pressure divergences" $\dot \theta$ diverges while $\theta$ is fine in `K' quantization or the `A' quantization with weak energy condition imposed or inverse triad modifications included.   As we will discuss in Sec. VII this indicates that these divergences will not amount to strong singularities. In fact we will find that the generic absence of strong singularities and boundedness of $\theta$, and correspondingly the existence of weak singularities and potential divergence of $\dot \theta$ are strongly related.

In summary, we find in this section that not all the quantities of interest may be bounded and finite even in effective dynamics. However, as we will see in proceeding sections on geodesics and strength of singularities that these divergences turn out to be harmless. We already get a hint of this in the fact that $\theta$ is fine at these events while $\dot \theta$ may diverge. The effective spacetime can be shown to be free of any strong singularities and that the geodesics do not break down at such events where these divergences in curvature invariants occur.

\section{Geodesic completeness}

The geodesic equation for the coordinate $x^i$ for a general metric can be written in the differential form as follows,
\be
(x^i)'' = \Gamma^{i}_{jk}(x^j)'(x^k)' \label{gdesics1}
\ee
where the prime denotes differentiation with respect to the affine parameter. Using the metric \eqref{metric}, we can find out all the Christoffel symbols and then integrate to get the geodesic equations for Bianchi-II metric. Using this procedure, the geodesic equations turn out to be,
\begin{align}
x'&= C_x \bigg(\frac{1}{a_1^2} + \alpha^2 z^2 \frac{1}{a_2^2} \bigg) + C_y \alpha z, \label{xdesic}\\
y'&= C_x \, \alpha \,z \, \frac{1}{a_2^2} + C_y, \label{ydesic}\\
z'&= -C_x\, \alpha \, y \, \frac{1}{a_3^2} + C_z \label{zdesic}\\
\intertext{and}
(t')^2 &= \varepsilon + \frac{C_x^2}{a_1^2}+ a_2^2 \bigg( C_y + C_x \alpha z \frac{1}{a_2^2} \bigg)^2 + a_3^2 \bigg( C_z - C_x\alpha y \frac{1}{a_3^2} \bigg)^2 . \label{tdesic}
\end{align}
Here $C_x, C_y$ and $C_z$ are constants of integration. The constant $\varepsilon$ is unity for timelike geodesics and is zero for null geodesics. The RHS of $(t')^2$ equation is non-negative, which means that $t'$ is well defined. The above set of four equations gives the velocity vector along the geodesic, which can be integrated to get the geodesic in parametric form.

From the geodesic equations we see that at least one of the velocity vectors will diverge when either of the scale factors vanishes. Indeed this is what happens in the cosmological singularity present 
in classical dynamics of Bianchi-II. However, this is not the case for LQC as we discuss below.

Even though there can be potential singularities in LQC, note that if triad variables do not vanish at a potential singularity geodesic equations do not break down. We have proved in the previous sections that the triad variables remain positive definite and finite for the dynamics in `A' quantization if either weak energy condition is satisfied or inverse triad corrections are included. In the `K' quantization this is achieved even without imposing any extra conditions. Hence, geodesic equations do not break down in the loop quantization of Bianchi-II spacetime. 

In all these scenarios the scale factors remain positive definite and finite for all finite evolution in terms of the coordinate time $t$. However, for co-moving observers the proper time is given by $t$ itself. This shows that the breakdown of the geodesics that happens in classical dynamics due to divergence in one or more of the scale factors, is avoided in effective dynamics for both `A' and `K' loop quantizations of the Bianchi-II spacetime.

Further note that the geodesic equations for the coordinates $y$ and $z$ are coupled to each other but are independent of the geodesic equation for coordinate $x$. These two equations \eqref{ydesic} and \eqref{zdesic} together form a linear system of first-order ordinary differential equations and hence have a unique global solution if initial positions are specified at a given initial time \cite{Vrabie2016}. 
 Since the RHS of these equations 
is well defined for all finite times $t$, the unique global solution is maximally extendable. Given that $y$ and $z$ can be maximally extended as functions of proper time, which really is $t$ for co-moving observers, then equation \eqref{xdesic} also has a maximally extendable solution because the RHS is of zeroth order in $x$ and only contains functions that are all finite for all finite time $t$. For the geodesic equation in time $t$, for comoving time the LHS turns out to be unity and the equation becomes a constraint between quantities on the RHS of \eqref{tdesic}, which are all finite in finite time $t$. This implies that the geodesics in case of Bianchi-II are maximally extendable in effective dynamics.\\

\noindent
\textbf{Remark 4:} In a recent paper, the authors showed that in a loop quantization of Kantowski-Sachs spacetime similar results on geodesics are obtained \cite{ks-strong}. Geodesics never break down in finite time evolution. Using the same arguments as in the previous paragraph its is straightforward to conclude the maximal extendibility of geodesics in loop quantum Kantowski-Sachs spacetime with arbitary matter. For the special case of a  vacuum Kantowski-Sachs spacetime, this result has been obtained recently \cite{william}.

\section{Lack of strong singularities}

The fact that geodesics do not break down at the points where potential divergences in curvature invariants occur tells us about the fate of point particles following the geodesics. To find what happens to an extended object as it falls into a region of strong curvature we have to look at how neighbouring geodesics are focused as we approach the potential singularity. For this purpose, analysis of strength of singularities comes in handy. A strong curvature singularity is characterized by the property that it crushes any in-falling object to zero volume regardless of what they are made of \cite{ellis1977singular,tipler1977singularities,ck-1985conditions}. A rigorous mathematical formulation along with the necessary conditions for the occurrence of a strong curvature singularity were first provided by Tipler \cite{tipler1977singularities}. However, we will use the necessary conditions provided by Clarke and Krolak \cite{ck-1985conditions} which generalize Tipler's conditions.

According to Clarke and Krolak \cite{ck-1985conditions}, if a singularity at affine parameter $\tau = \tau_o$ is a strong curvature singularity, then, for a timelike (or null) geodesic running into the singularity, the integral
	\begin{equation}
	K^{i}_{j} =\int_0^{\tau} \mathrm{d} \tilde\tau |R^{i}_{4j4} (\tilde\tau)| \label{krolak}
	\end{equation}
	does not converge as $\tau \rightarrow \tau_{o} $. Else the singularity is weak. 
	First of all note that for co-moving observers, the coordinate time $t$ is the affine parameter itself. Hence,  to check whether these conditions are satisfied by the spacetime under consideration, we must look at the evolution of the Riemann tensor components as a function of the coordinate time $t$.

As mentioned in the section on curvature invariants, the Riemann curvature tensor is obtained from the second derivative of the metric. The metric is a function of time through the triad variables, and also depends on the coordinate $z$. Hence any term obtained via the second derivative of the metric will necessarily be one of the following three types, i.e. the following three possibilities exhaust the types of terms that can appear in the Riemann tensor components:\\
\begin{enumerate}[T-I.]
\item Terms of type $f(p_1,p_2,p_3,z)$.
\item Terms of type $\bigg(\frac{\dot p_1}{p_1}\bigg)^m \bigg(\frac{\dot p_2}{p_2}\bigg)^n \bigg(\frac{\dot p_3}{p_3}\bigg)^q f(p_1,p_2,p_3,z)$ where $m,n,q$ are positive integers and $m+n+q=2$.\\
\item Terms of type $\frac{\ddot p_i}{p_i}f(p_1,p_2,p_3,z)$, where $i$ can be $1,2$ or $3$.\\
\end{enumerate}
The components of the curvature tensors are made of sums or differences of these three types of terms. Using the fact that,
\begin{equation}
\int_a^b |f_1 + f_2 + ... + f_n| dt \leq \int_a^b |f_1|dt + \int_a^b |f_2|dt + ..... + \int_a^b |f_n| dt
\end{equation}
we only need to consider divergence properties of the three types of terms individually to conclude about the divergence of the integral in \eqref{krolak}.

We have shown in Sec. III that in the `A' quantization, both $p_i$ and $\frac{\dot p_i}{p_i}$ will be finite for all finite times if weak energy condition is imposed or inverse triad corrections are incorporated. In the `K' quantization this is the case without including either weak energy condition or the inverse triad corrections. Hence the terms of type T-I and T-II will remain finite for any finite time $t$. Consequently their integral over a finite range of $t$ will also be finite. Hence, we conclude that the terms of type T-I and T-II in the integral \eqref{krolak} do not diverge. The same can not be said about the terms of the type T-III because they contain second derivatives of triads. As noted in Sec. V these can potentially diverge for events where pressure or its derivatives become infinite at finite energy density. Thus, only terms of type T-III could possibly lead to divergence of the integral in eq.(\ref{krolak}). {{In fact, these are the very scenarios which make the curvature invariants and $\dot \theta$ diverge.}} However as we show below, these problematic second derivatives of the triad variables are removed by integration. Let us consider such a term for the integral in \eqref{krolak}. Then it can be expressed as
\be
\int_0^{\tau_o} \frac{\ddot p_i}{p_i}f(p_1,p_2,p_3)\mathrm{d}\tau  = \frac{\dot p_i}{p_i}f(p_1,p_2,p_3) \bigg\vert_0^{\tau_o} - \int_0^{\tau_o} \dot p_i \bigg(\frac{\mathrm{d}}{\mathrm{d}\tau}f_1(p_1,p_2,p_3)\bigg)\mathrm{d}\tau ~.
\ee
We note that both the terms on the RHS are finite. Hence, the terms of type T-III do not lead to divergence of the integral \eqref{krolak}. Consequently the integral \eqref{krolak} does not diverge in the effective dynamics of LQC in Bianchi-II spacetime. So we have proved that the necessary conditions for existence of strong singularities are not satisfied by the effective dynamics of the `K' quantization or `A' quantization with weak energy condition or inverse triad corrections. In summary, the effective spacetime in `A' (with either weak energy condition or inverse triad modifications included) and `K' quantizations is free of strong singularities. Notably the events where potential divergences in curvature invariants or $\dot \theta$ occur correspond to weak curvature singularities.\\

\noindent
{\bf{Remark 5:}} In the case of `A' quantization when neither weak energy condition nor inverse triad modifications are included, $p_1$ can vanish in finite time and $\dot p_1/p_1$ is not bounded. In this case, all the terms: T-I, T-II and T-III result in divergences in eq.\eqref{krolak}. The cigar singularity in this quantization thus turns out to be a strong curvature singularity. It is straightforward to see from our arguments in the previous section that the geodesics also break down in this case.

\section{Conclusions}
There is a strong and growing evidence that quantum gravitational effects as understood in loop quantum gravity can yield generic resolution of singularities. 
The complete picture from the loop quantum gravity is missing but loop quantization of several symmetry reduced spacetimes now exists \cite{as-status}. 
A careful study of these spacetimes in context of understanding geodesic completeness and resolution of strong curvature singularities is expected to provide invaluable hints 
to the existence of a non-singularity theorem in quantum gravity. This work is a part of a systematic investigation of various loop quantum spacetimes and the generic resolution of strong curvature singularities 
due to underlying quantum geometry effects for arbitrary matter. Starting from the isotropic and homogeneous cosmologies for the spatially flat model for the first time geodesic completeness and 
resolution of all strong curvature singularities were demonstarted in Ref. \cite{ps09}. This work also established that weak curvature singularities can exist in LQC. These results were confirmed in the 
spatially closed models in LQC using various phenomenological equations of state \cite{psfv}. A generalization to the Bianchi-I spacetimes was performed in Ref. \cite{ps11}, and recently these results have been extended to the loop quantization of Kantowski-Sachs spacetime with arbitrary matter \cite{ks-strong}.

In continuation of the above works on the geodesic completeness and the resolution of strong singularities in effective dynamics of LQC, our goal in this manuscript was to generalize these results to the case of loop quantum Bianchi-II spacetimes. Bianchi-II is an important spacetime to study to understand generic approach to singularities and their resolution. Various studies in classical GR on Mixmaster behavior and BKL conjecture have shown that in a generic approach towards singularity captured by Bianchi-IX spacetime, the spacetime asymptotically can be modeled as undergoing long phases that mimic Bianchi-I spacetime mediated by short transition periods where the spacetime resembles Bianchi-II spacetime \cite{Berger1998}. In this manuscript, we have investigated the resolution of singularities in Bianchi-II spacetime in the effective dynamics of LQC using two different quantization prescriptions - the connection operator based  `A' quantization and the extrinsic curvature based `K' quantization. In the loop quantization of Bianchi-II spacetime, unlike in the Bianchi-I spacetime it is not possible to express field strength of the connection in terms of holonomies over the closed loop which yield an algebra of almost periodic functions of connection. The strategy then is to introduce a connection operator defined using open loop segments and express quantum constraint in terms of the connection operator. This leads to the `A' quantization \cite{awe-bianchi2}. However, in the Bianchi-II spacetime one can also perform a gauge fixing and treat extrinsic curvature as connection. The resulting quantum constraint leads to `K' quantization. The latter has been obtained for Bianchi-IX spacetime in Ref. \cite{pswe}, where its merits in contrast to `A' quantization were also noted for the first time. In particular, it was shown that unlike `A' quantization of Bianchi-IX spacetime, the `K' quantization results in generically bounded expansion and shear scalars. In this manuscript, we considered the simpler case of this quantization for Bianchi-II model and studied its effective dynamics. 

The results from `A' and `K' loop quantizations of the Bianchi-II spacetime exhibit some striking differences and similarities at the same time. A key difference is that if inverse triad modifications are not included in the Hamiltonian constraint or weak energy condition is not imposed on the matter content then the conenction operator based `A' quantization does not lead to a generic resolution of strong singularities. We show that in this case even though energy density is bounded from above, it goes to negative infinity at the potential cigar type singularities that may still occur. And the directional Hubble rates, the expansion and shear scalars may still diverge in finite time. We note that the potential divergence of expansion and shear scalar in this case was earlier noted in Ref. \cite{bgps-spatial}. In particular, we show that a cigar like singularity can arise in this quantization. Such a singularity not only results in divergences in curvature invariants but also breakdown of geodesic evolution. The analysis of strength of singularity shows that the singularity is strong curvature type. Since the expansion scalar is unbounded in this case, the geodesic equations also break down in finite time. 

In contrast, for the `K' quantization the situation is very similar to the the case of isotropic models, the Bianchi-I and the Kantowski-Sachs spacetime in LQC. First, we proved that all triads neither vanish nor diverge in a finite time evolution. The Hubble rates, expansion scalar and the shear scalar are universally bounded. The curvature invariants can diverge but only if there are divergences in pressure or its derivatives at a finite energy density -- a situation which arises only for highly exotic equations of state of matter. Such potential divergences turn out to be weak curvature singularities. An important result which we prove in this setting is that {\it{all}} strong curvature singularities are generically resolved. Examination of geodesics in this case reveals that the effective spacetime is geodesically complete. The geodesics are maximally extendible beyond potential curvature divergence events.

The situation improves drastically for the `A' quantization if in the effective Hamiltonian inverse triad modifications are included. These are often excluded from the discussion of effective dynamics even though they are in general one of the two ways quantum geometric effects manifest themselves. The importance of inclusion of inverse triad modifications has been lately emphasized in some works \cite{bgps-spatial,corichi-bianchi3}. Further, inverse triad modifications exclusively result in interesting phenomenological effects (see for eg. \cite{spatial}) and even bounce without holonomy modifications in spatially closed model \cite{st}. The effective dynamics of the Bianchi-II model in `A' quantization in the presence of inverse triad modifications is such that all the triads remain non-vanishing and finite for all finite time, and the directional Hubble rates and expansion and shear scalars do not diverge. The curvature invariants can still diverge in this case, exactly as in the `K' quantization, but all such events are harmless. We show that geodesics do not break down and they are not strong curvature singularities. 

We find that in the case of Bianchi-II spacetime with `A' quantization there is another way to achieve same results without using inverse triad modifications. It is if 
weak energy condition is imposed on the matter content. This results in binding the energy density from below as well and then one obtains a 
 non-singular evolution of the dynamics for all finite times. If weak energy condition is imposed then  the effective spacetime is geodesically complete and no strong curvature cosmological singularity is reached in finite time. As in the case of `K' quantization, curvature invariants such as Ricci and Kretschmann scalar may still diverge but such divergences are not associated to a strong singularity. 
 It is important to note that this is for the first time in LQC where singularity resolution gets tied to the energy conditions in a quantization. It will be worth investigating further if this turns out to be a 
 more common feature in loop quantization of more complex spacetimes. Notably, singularity theorems in GR rely strongly on energy conditions. Whether or not a `non-singularity theorem' in quantum gravity 
 would depend on energy conditions is an open issue. 
 
Let us finally note that in the loop quantization of all other spacetimes neither inverse triad modifications nor weak energy condition were ever needed to obtain generic resolution of singularities \cite{ps09,ps11,ks-strong}. 
Holonomy modifications were always sufficient to achieve resolution of all strong curvature singularities and geodesic completeness in the effective spacetime. This is ture even in the case of loop quantization of Kantowski-Sachs spacetime where as in Bianchi-II spacetime, spatial curvature is non-vanishing \cite{ks-strong}.  This is no longer the case for the loop quantization of Bianchi-II spacetime in case of connection operator based method -- the `A' quantization. Holonomy modifications are insufficient to avoid strong singularities. One must use inverse triad modifications or weak energy condition to rescue from strong curvature singularity catastrophe. Needless to say, the role of both of these inputs is extremely important to be further understood in more general loop quantizations. In a sharp contrast, we must note the relatively easy way the `K' quantization achieved the generic singularity resolution and geodesic completeness in the Bianchi-II spacetime. 
This is a clear evidence that `K' quantization which turns out to be equivalent to `A' quantization in absence of intrinsic curvature yields generic resolution of singularities comparatively  easily. In this sense, it can be regarded 
as a favored choice. It remains to be seen whether loop quantization of more general spacetimes validates this assertion.

\begin{acknowledgments}
We are grateful to William Cuervo for insightful discussions. 
This work is supported by NSF grants PHY-1404240 and PHY-1454832. We thank an anonymous referee for her/his valuable comments which improved the presentation in our manuscript. 
\end{acknowledgments}

\end{document}